\documentclass[letterpaper, 10 pt, conference]{IEEEtran}  

\IEEEoverridecommandlockouts  

\usepackage{cite}
\usepackage{comment}
\usepackage{mathtools}
\usepackage{bm}
\usepackage{amsmath, amssymb, amsfonts}
\usepackage{algorithm}
\usepackage{algpseudocode}
\usepackage{graphicx}
\usepackage{textcomp}
\usepackage{stfloats}
\usepackage{xcolor}
\newcommand{\cmmnt}[1]{}
\def\BibTeX{{\rm B\kern-.05em{\sc i\kern-.025em b}\kern-.08em
    T\kern-.1667em\lower.7ex\hbox{E}\kern-.125emX}}
    
\begin{document}


\title{Cloud-Based Scheduling Mechanism for Scalable and Resource-Efficient Centralized Controllers}


\author{Achilleas Santi Seisa$^{*}$, Sumeet Gajanan Satpute and George Nikolakopoulos
\thanks{This project has received funding from the European Union’s Horizon 2020 research and innovation programme under the Marie Skłodowska-Curie grant agreement No 953454.}
\thanks{The authors are with the Robotics and AI Group, Department of Computer, Electrical and Space Engineering, Lule\aa\,\, University of Technology, Lule\aa\,\,}
\thanks{$^{*}$Corresponding Author's email: {\tt\small achsei@ltu.se}}
}

\maketitle


\begin{abstract}
%

This paper proposes a novel approach to address the challenges of deploying complex robotic software in large-scale systems, i.e., Centralized Nonlinear Model Predictive Controllers (CNMPCs) for multi-agent systems. The proposed approach is based on a Kubernetes-based scheduling mechanism designed to monitor and optimize the operation of CNMPCs, while addressing the scalability limitation of centralized control schemes. By leveraging a cluster in a real-time cloud environment, the proposed mechanism effectively offloads the computational burden of CNMPCs. Through experiments, we have demonstrated the effectiveness and performance of our system, especially in scenarios where the number of robots is subject to change. Our work contributes to the advancement of cloud-based control strategies and lays the foundation for enhanced performance in cloud-controlled robotic systems.

\end{abstract}


\section{Introduction}
\label{sec:intro}
Cloud robotics has been a topic of discussion for several years~\cite{robotics7030047, hu2012cloud}, but the widespread utilization of cloud computing in robotic systems has remained limited to specific applications, such as Computer Vision (CV)~\cite{tian2019fog, beigi2017real}, learning (machine 
learning, robot learning, etc.)~\cite{gao2018novel, nakanoya2023co, penmetcha2021deep}, and Simultaneous Localization And Mapping (SLAM)~\cite{ahmed2023real}\cmmnt{\cite{benavidez2015cloud, sarker2019offloading, riazuelo2015roboearth}}. However, the potential advantages of cloud computing extend even to more traditional robotics concepts that require significant computational resources \cmmnt{to enhance system performance}. One such concept is the trajectory tracking control with embedded collision avoidance, such as Nonlinear Model Predictive Control (NMPC)~\cite{santos2023nonlinear}, which can greatly benefit from leveraging cloud computing capabilities.

\cmmnt{By harnessing cloud computing, trajectory control modules can offload intense computation to powerful cloud servers, alleviating the burden on onboard hardware, and enabling more efficient and sophisticated control strategies. This approach opens up new possibilities for real-time optimization and enhanced performance in robotic systems, ultimately advancing the capabilities and applications of cloud-based robotics.}

In~\cite{zhang2023kuberos}, a \cmmnt{comprehensive}framework called KubeROS is introduced to tackle the challenges of deploying complex robotic software in large-scale systems.\cmmnt{Although this proposed work also falls within the realm of cloud robotics and addresses the deployment of intricate robotic software in large-scale settings, there are distinct differences that set our approach apart.} Unlike KubeROS, our proposed system does not require robots to be part of the Kubernetes cluster. This provides the advantage of flexibility, allowing the number of robots to dynamically change\cmmnt{as needed}. Additionally, our system introduces a fully dynamic and automated solution to overcome the scalability limitation of centralized control schemes, achieved through the implementation of\cmmnt{an intelligent} a scheduling mechanism. \cmmnt{By drawing attention to these specific differences, we aim to highlight the unique contributions of our proposed system while contextualizing them within the broader landscape of cloud robotics and large-scale robotic software deployment.}

The pursuit of efficient resource management in cloud-controlled robotic systems presents unique challenges, particularly when employing computationally intensive control methods, such as NMPC. The complexity and computational demands of centralized scenarios, where a single controller governs the trajectories of multiple agents (Centralized NMPC i.e., CNMPC), further exacerbate these challenges. The performance of these controllers is influenced by various factors, including the number of agents, the parameters of the NMPC (e.g., prediction horizon, optimization solver iterations, etc.), and the constraints of the system~\cite{seisa2022edge}.

The debate between distributed and centralized control has been a topic of extensive research. Previous studies, such as~\cite{bertilsson2022centralized} and~\cite{hu2018centralize}, have analyzed centralized and distributed control schemes for swarms in detail. These works concluded that while centralized approaches clearly outperform distributed ones in terms of performance, the scalability limitations of centralized schemes render them unsuitable for large-scale multi-agent systems. In light of this challenge, the present article proposes a novel approach to overcome the scalability limitation of centralized control schemes. \cmmnt{Some previous works have suggested switching between distributed and centralized Model Predictive Control (MPC) approaches for multi-agent systems, as in~\cite{mansouri2015distributed, wehbeh2020distributed, zhan2022data, sahu2021model}, while other works have presented standalone centralized MPC schemes for multi-agent systems~\cite{lindqvist2020collision, seisa2022cnmpc}. In contrast, the proposed approach}We introduce a scalable framework based on a scheduling mechanism that can generate multiple centralized control schemes, thus retaining the high-performance characteristic of centralized control while addressing scalability concerns.

\cmmnt{
To enable real-time optimization and expedite solution generation, our objective is to dynamically monitor system resources within the centralized controller. This necessitates a responsive and adaptable framework capable of allocating resources based on the number of agents and CNMPC parameters. Leveraging the capabilities of a cutting-edge real-time cloud environment, we can harness the required resources seamlessly. Facilitated by a robust Kubernetes architecture, it is ensured that the CNMPC runs on a designated worker node, equipped with the specified computational resources.
}
\begin{figure}
    \centering
    \includegraphics[width=0.88\linewidth]{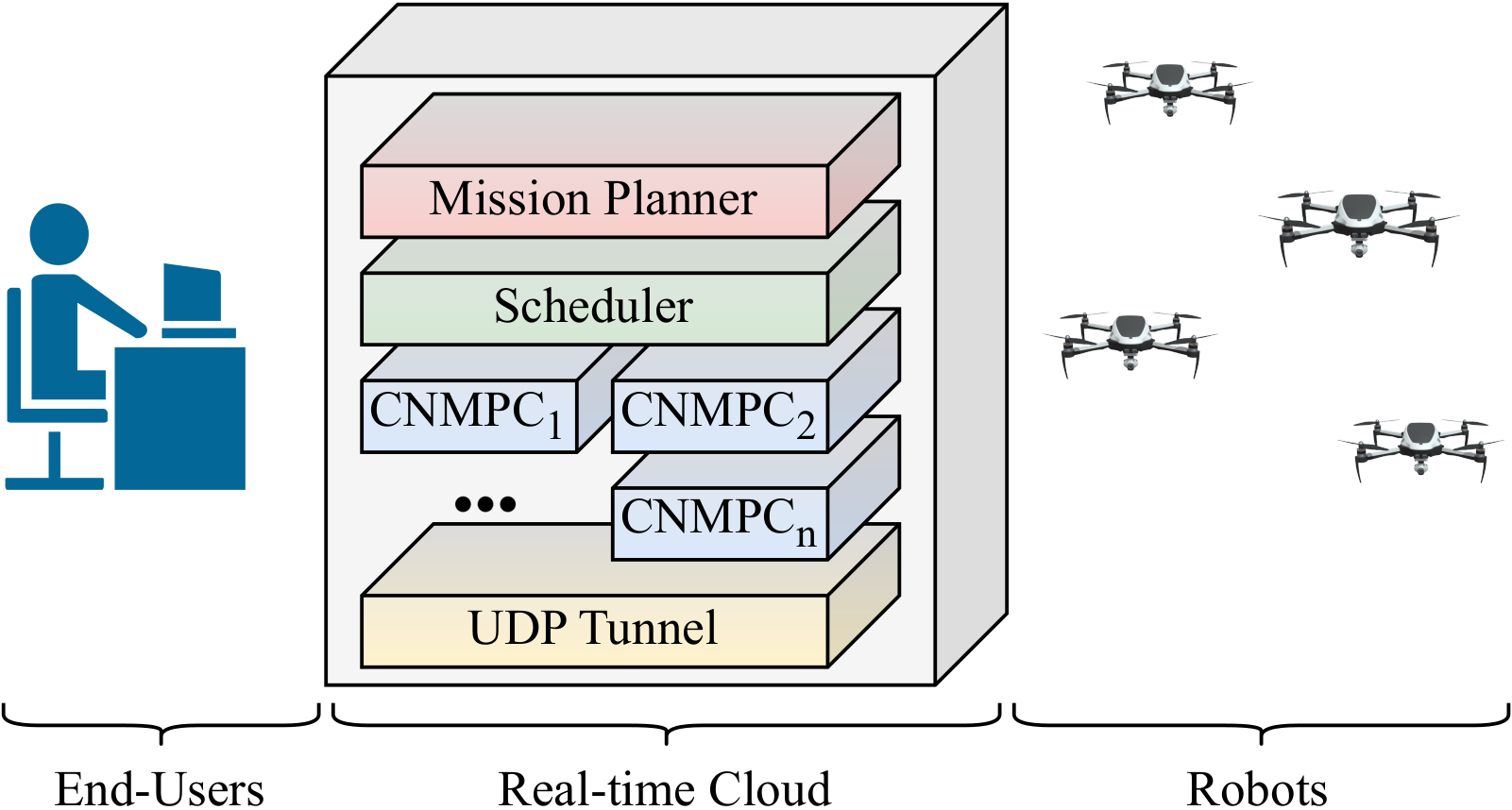}
    \caption{High-level overview of the proposed framework in the real-time cloud with its core components.}
    \label{fig:concept}
    \vspace{-0.8cm}
\end{figure}
In this context, we present the development of a sophisticated scheduler operating at a frequency of 1Hz. Operating from a mission planning node, the scheduler meticulously monitors the number of agents and the CNMPC parameters. Any changes detected in these parameters trigger the deployment of a new CNMPC or the update of the existing one. Consequently, centralized controllers are efficiently allocated to worker nodes that can precisely fulfill the requested computational resources, promoting optimal performance and responsiveness, and communicating with the external world with a data transmission module. The realization of this framework is achieved with the development of these main modules in the real-time cloud, as depicted in Fig.~\ref{fig:concept}.

\cmmnt{By dynamically monitoring resources in a real-time cloud environment and by integrating Kubernetes-based scheduling mechanisms, our approach offers a sophisticated solution to overcome the computational challenges associated with CNMPC. This research contributes to the advancement of resource-aware control strategies and lays the foundation for enhanced performance in cloud-controlled robotic systems.}

The proposed scheduling mechanism overcomes the limitations of centralized control schemes by enabling the dynamic allocation of agents and CNMPC parameters for each CNMPC. This flexibility allows for the efficient control of varying numbers of agents, making it suitable for applications with changing agent counts in comparison to~\cite{lindqvist2020collision}, where the number of agents has to be predefined. By leveraging the Kubernetes cluster, the proposed mechanism can handle a large number of agents. The system achieves this by deploying the corresponding centralized controllers and allocating them to worker nodes equipped with appropriate resources, ensuring efficient control over numerous agents, thus enabling the application of CNMPCs in complex and large-scale systems.

\cmmnt{The cloud environment plays a crucial role in managing the allocation of computational resources. By monitoring and controlling the utilization of resources by each controller, the proposed system ensures the efficient and appropriate utilization of Central Processing Unit (CPU) cores and memory, optimizing resource allocation and minimizing wastage. To handle potential challenges such as communication issues or cloud unavailability, the proposed system incorporates safety fallback actions for the agents (discussed in our previous work~\cite{10228114}). These fallback actions provide a safety net to ensure that the agents can continue functioning even in adverse conditions, maintaining system stability and robustness. }

In summary, the contributions of this work encompass the novel establishment of a cloud framework for multi-agent closed-loop robotic applications with the ability to handle scalability and flexibility through dynamic allocation of computational resources. Compared to~\cite{dhiyanesh2012dynamic}, which primarily focuses on dynamic resource allocation through decisions about execution location\cmmnt{(onboard, cloud, robot collaborative execution, or hybrid)}, our work maximizes the utilization of cloud resources by dynamically allocating and continuously monitoring application resources within the cloud infrastructure. Additionally, the proposed framework is capable of handling a large number of agents and sequentially providing optimized resource allocation. \cmmnt{Finally, we introduce the incorporation of safety fallback actions for enhanced system robustness. These contributions advance the field of cloud-based control systems, offering practical solutions for efficient and reliable control of CNMPCs in various applications.}

\begin{figure*}
    \centering
    \includegraphics[width=\textwidth]{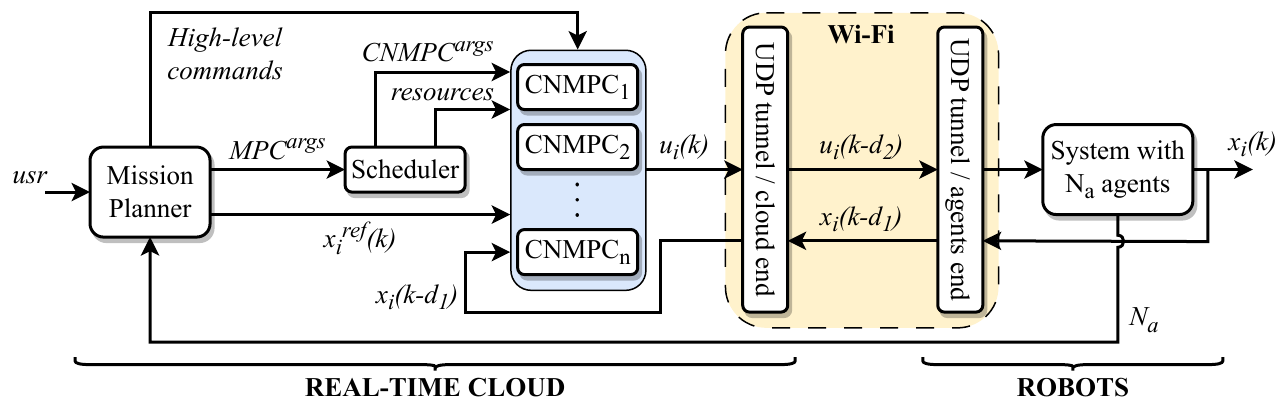}
    \caption{Overview of the block diagram. Real-time cloud includes the mission planner, the scheduler, the controllers, and the proxy server (UDP tunnel). Robots include the multi-agent system and the agent end of the UPD tunnel.}
    \label{fig:cnmpc}
\end{figure*}


\section{Scheduling Mechanism}
\label{sec:scheduling_mechanism}
The scheduling mechanism serves as a fundamental component of this work as it not only monitors the available resources but also facilitates the deployment of CNMPCs.\cmmnt{ This section presents the four modules of the mechanism.}

\subsection{Mission Planner}
\label{sec:planner}
To effectively coordinate the operations of the agents and determine the required parameters for the CNMPCs, a mission planner has been developed. \cmmnt{The planner receives inputs either directly from the user or from predefined mission templates, such as multi-agent exploration, inspection, or logistic-robot tasks. In the scope of this work, }The mission planner receives the desired number of cloud-controlled agents, denoted as $agent_{num}^{d}$, along with high-level commands including take-off and safety-land. Subsequently, the mission planner publishes this information to other relevant nodes, such as the scheduler, and the CNMPCs, aiming to form units composed of agents within the same CNMPC. Additionally, it provides the desired trajectories, represented by $x_{i}^{ref}$, for all the cloud-controlled agents involved in the mission, as depicted in the block diagram of Fig.~\ref{fig:cnmpc}. 

\subsection{Scheduler}
\label{sec:scheduler}
The scheduler is the most crucial part of the scheduling mechanism, serving as its core component. It receives the necessary information from the mission planner, including the number of agents and the CNMPC parameters ($CNMPC^{args}$). Leveraging this information, the scheduler dynamically generates the necessary deployments required for the operation of the CNMPCs, considering as well the computational requirements. In addition to creating deployments, corresponding services are established that facilitate seamless message exchange between the cloud and the agents.
\cmmnt{The scheduler leverages various Kubernetes features via the Kubernetes command-line tool, kubectl. As a result, }The scheduler not only monitors the deployment of controllers and their resource utilization but also ensures continuous execution by creating replicas. This redundancy guarantees that cloud controllers remain operational, even in cases where pods running controllers face interruptions in their execution. To enhance the scheduler's effectiveness, insights are extracted from the Kubernetes environment. These logs provide essential feedback, allowing the scheduler to fine-tune its performance and optimize the allocation of computational resources. Overall, the scheduler\cmmnt{ serves as a pivotal component within the cloud-based control system,} harnessing Kubernetes capabilities to facilitate dynamic deployments, resource management, and robust communication between cloud controllers and agents.

\cmmnt{The monitoring of the system by the scheduler is established mainly based on the resource requirements and the CNMPC parameters or $CNMPC^{args}$, as shown in Fig.~\ref{fig:cnmpc} and described in the following subsections. The scheduler forwards corresponding information to each controller, such as the prediction horizon, and decides how many agents should each CNMPC control (with $agent_{max}$ the maximum controllable agent per CNMPC due to the optimizer's limitations) and how many resources it will need. }

\subsubsection{Monitoring the Deployment of CNMPCs}
\label{sec:cnmpc_monitoring}
This task required a comprehensive understanding of both the mission planner and the intricacies associated with the function of CNMPCs. In particular, it was crucial to determine the maximum number of agents that can be effectively controlled by a single CNMPC, thereby facilitating the calculation of the required number of CNMPCs to govern all available agents. This calculation was performed empirically, and the corresponding $agent_{max}$ value was selected. Based on this knowledge, a dynamic deployment strategy for CNMPCs was devised, as illustrated in Algorithm~\ref{alg:cnmpc_deplyment}. While~\cite{zhang2023kuberos} uses a load balancer to manage the number of Virtual Machines (VMs) required for motion planning, our approach takes a more dynamic and flexible approach by allowing the scheduler to generate and terminate deployments as needed based on various inputs such as the number of agents and mission planner objectives.

\begin{algorithm}
\caption{Deployment of CNMPCs based on the desired number of agents ($agent_{num}^{d}$)}\label{alg:cnmpc_deplyment}
\begin{algorithmic}[1]
    \If{$agent_{num}^{d} \neq agent_{num}^{old}$}
        \State $CNMPC_{num} = int(\frac{agent_{num}^{d}-1}{agent_{max}}+1)$
        \If{$agent_{num}^{d} < agent_{num}^{old}$}
            \For{$j = CNMPC_{num} : CNMPC_{num}^{old}$}
                \State $\text{There are unnecessary deployments}$
                \State $\text{Delete unnecessary deployments}$
            \EndFor
            \For{$j = 2 * agent_{num}^{d} + 1 : 2 * agent_{num}^{old} + 1$}
                \State $\text{There are unnecessary services}$
                \State $\text{Delete services}$
            \EndFor
        \EndIf
        \If{$agent_{num}^{d} \neq 0$} 
            \State $agent_{CNMPC} = int(\frac{agent_{num}^{d}-1}{CNMPC_{num}})$
            \State $agent_{CNMPC}^{float} = \frac{agent_{num}^{d}-1}{CNMPC_{num}}$
            \State $counter = 0$        
            \For{$j = 0 : CNMPC_{num}$}
                \If{$agent_{CNMPC} = agent_{CNMPC}^{float}$}
                    \State $agent_{CNMPC} = int(\frac{agent_{num}^{d}-1}{CNMPC_{num}})$
                    \State $\text{Create or update deployments}$
                    \State $\text{Create or update services}$
                \Else
                    \State $counter = counter +1$
                    \State $agent_{CNMPC} = int(\frac{agent_{num}^{d}-1}{CNMPC_{num}}) + 1$
                    \State $agent_{CNMPC}^{float} = \frac{agent_{num}^{d} - counter}{CNMPC_{num}}$
                    \State $\text{Create or update deployments}$
                    \State $\text{Create or update services}$
                    \State $agent_{CNMPC} = agent_{CNMPC} - 1$
                \EndIf
            \EndFor
        \EndIf
        \State $agent_{num}^{old} = agent_{num}$
        \State $CNMPC_{num}^{old} = CNMPC_{num}$
    \EndIf
\end{algorithmic}    
\end{algorithm}
\setlength{\textfloatsep}{0pt}

The total number of CNMPCs to control all the agents is denoted as $CNMPC_{num}$, while $agent_{num}^{old}$ and $CNMPC_{num}^{old}$ describe the number of agents and the number of CNMPCs in the previous iteration, respectively. The parameters $agent_{CNMPC}\in\mathbb{Z}^{+}$ and $agent_{CNMPC}^{float}\in\mathbb{R}^{+}$ are describing the number of agents per CNMPC. Finally, $counter$ is an auxiliary variable that is needed for the correct deployment of the controllers.

\subsubsection{Resource Allocation}
\label{sec:resource_allocation}
In comparison to other cloud or edge architectures for robot control as in our previous work~\cite{seisa2022cnmpc}, the current work takes into account the computational requirements of the application to strategically deploy CNMPCs to worker nodes, equipped with the necessary resources. Opposed to~\cite{zhang2023kuberos}, where the resources are managed at a high level based on the available hardware, we experimentally analyzed and estimated the computational effort exerted by the CNMPCs, considering factors, such as the number of agents and the CNMPC parameters. This analysis leads us to formulate Eq.~\eqref{eq:resources_cpu}, and~\eqref{eq:resources_mem}:
\begin{subequations}\label{eq:resources_cpu}
\begin{eqnarray}
        CPU^{d}_{min} = f_1(x) = a * \frac{x - 1}{1 - N} + CPU^{n}_{min} & (cores) \hspace{2mm}\label{eq:cpu_min} \\
        CPU^{d}_{max} = f_2(x) = a * \frac{x - 1}{1 - N} + CPU^{n}_{max} & (cores) \hspace{2mm}\label{eq:cpu_max}
\end{eqnarray}
\end{subequations}
\vspace{-4mm}
\begin{subequations}\label{eq:resources_mem}
\begin{eqnarray}
        M^{d}_{min} = g_1(x) = b * \frac{x - 1}{1 - N} + M^{n}_{min} & (MiB)\hspace{2mm}\label{eq:mem_min} \\
        M^{d}_{max} = g_2(x) = b * \frac{x - 1}{1 - N} + M^{n}_{min} & (MiB)\hspace{2mm}\label{eq:mem_max}
\end{eqnarray}
\end{subequations}
The parameter $x\hspace{-2mm}\in\hspace{-2mm}\mathbb{R}$ describes the number of agents for each CNMPC ($agent_{CNMPC}$), $N\hspace{-2mm}\in\hspace{-2mm}\mathbb{R}$, where $0\hspace{-1mm}\leq\hspace{-1mm}N\hspace{-1mm}<\hspace{-1mm}1$, corresponds to the CNMPC prediction horizon and rate parameters ($CNMPC^{args}$), while $a,b\hspace{-0.5mm}\in\hspace{-0.5mm}\mathbb{Z}^+$ are known scalars. Utilizing the derived equation, we establish CPU and memory (M) allocations for each CNMPC within a specified minimum and maximum range $[CPU^{d}_{min}, CPU^{d}_{max}]$ and $[M^{d}_{min}, M^{d}_{max}]$, respectively. In addition, the minimum and maximum CPU and memory requirements for CNMPC execution are defined as $CPU^{n}_{min}, CPU^{n}_{max}, M^{n}_{min}$, and $M^{n}_{max}\hspace{-0.5mm}\in\hspace{-0.5mm}\mathbb{Z}^+$, respectively, based on the minimum resource requirement and the maximum available resources without depleting them. This ensures that each CNMPC efficiently utilizes the appropriate number of CPU cores and memory, thereby optimizing overall efficiency, while avoiding unnecessary overhead.
\subsection{Data Flow}
\label{sec:data_flow}
The data flow in our cloud-based control system facilitates seamless communication and coordination between the cloud-controlled agents and the real-time cloud. It operates at two levels and leverages a proxy server with a User Datagram Protocol (UDP) tunnel for communication between agents and the cloud, and is illustrated within the yellow area in Fig.~\ref{fig:cnmpc}. Additionally, communication within the Kubernetes cluster relies on the Robotic Operating System (ROS) networking mechanism. In comparison to traditional distributed systems, where agents often require direct communication with one another, the proposed approach streamlines communication. All agents communicate exclusively with the cloud, eliminating the need for direct peer-to-peer interactions. The cloud provides all the necessary information for the agents to operate and interact with their environment. This streamlined communication paradigm enhances system efficiency and reduces the complexity of agent-to-agent communication.

\subsubsection{Data Transmission Through a Proxy Server}
\label{sec:udp_tunnel}
For the transmission of ROS messages between the agents and the real-time cloud, a proxy server is utilized with a UDP tunnel comprising client and server nodes. One end of the tunnel is running at each agent, while the other end is running at the proxy server. The agents send their positional information, including position, velocity, and orientation ($x_{i}(k)=[p_i(k),\dot{p}_i(k),q_i(k)]^T$, where $i=1,\dots,agent_{num}$), to the cloud using the proxy server's Internet Protocol (IP) address. Before transmission, ROS messages are transformed into byte arrays to facilitate data transfer. The proxy server extracts this information from the byte arrays and forwards it to the ROS nodes running within the Kubernetes pods as ROS messages. This information arrives in the cloud with uplink delay denoted with $d_{1}$, thus the positional information is described as $x_{i}(k-d_{1})$. Similarly, control actions, such as roll, pitch, yaw, and thrust ($u_{i}(k-d_{2})$, where $d_{2}$ describes the downlink delays), are transformed into byte arrays and sent to the agents from the cloud through the proxy server. To ensure smooth and efficient transmission, sockets are dynamically generated, providing specified ports for each message's exchange. Given that CNMPCs are stateful applications, sharing information across all relevant nodes within the Kubernetes cluster is crucial. By employing a proxy server, which holds all necessary information and utilizes ROS for internal cluster communication, we effectively address communication challenges and facilitate seamless migration between CNMPC applications.
\subsubsection{Robotic Operating System}
\label{sec:ros}
Within the Kubernetes cluster, all pods are part of the same network, enabling ROS nodes to communicate freely with each other using ROS subscribing and publishing mechanisms. To facilitate communication, all ROS nodes must register with the same ROS master, which runs independently in its own Kubernetes pod to prevent potential interference during execution.
\subsection{Centralized Nonlinear Model Predictive Controllers}
\label{sec:cnmpc}
\subsubsection{Robot kinematic model}
\label{sec:model}
The utilization of MPC for trajectory control of Unmanned Aerial Vehicles (UAVs) has been studied extensively in previous works~\cite{santos2023nonlinear, okasha2022design, islam2019comparative}. In this work, the CNMPC utilizes the UAV model described in~\cite{small2019aerial}, and is based on NMPCs that can compensate for latency as described in~\cite{sankaranarayanan2023paced}. The UAVs are represented as fixed-body, six degrees-of-freedom robots, as in Eq.~\eqref{eq:kinematics}:
\begin{subequations}
\label{eq:kinematics}
    \begin{align}
        \mathbf{\Ddot{p}_i}(t) &= \frac{1}{m} \mathbf{R_i}(\mathbf{q_i}(t)) \mathbf{F_i}(t-\tau) + \mathbf{G} - \mathbf{A}\mathbf{\dot{p}_i}(t) \label{eq:pos_dyn} \\
        \dot{q}_{\phi,i}(t) &= \frac{1}{\alpha_\phi}(K_\phi q_{\phi,i}^d(t - \tau) - q_{\phi,i}(t)) \label{eq:phi_dyn} \\
        \dot{q}_{\theta,i}(t) &= \frac{1}{\alpha_\theta}(K_\theta q_{\theta,i}^d(t - \tau) - q_{\theta,i}(t)) \label{eq:theta_dyn}
    \end{align}
\end{subequations}
The parameters $\mathbf{p_i} \triangleq \begin{bmatrix} p_{x,i}(t) & p_{y,i}(t) & p_{z,i}(t)\end{bmatrix}^T \in \mathbb{R}^3$ and $\mathbf{q_i} \triangleq \begin{bmatrix} q_{\phi,i}(t), q_{\theta,i}(t), q_{\psi,i}(t) \end{bmatrix}^T \in \mathbb{R}^3$ describe the position and orientation of each UAV, where $i=1,\dots,agent_{num}$, while $m$ is the mass of the UAVs, $\mathbf{R_i} \in \mathbb{R}^{3 \times 3}$ is the Euler angle rotation matrices, and $\mathbf{F_i}(t-\tau)\triangleq \begin{bmatrix} 0 & 0 & F_{z,i}(t-\tau) \end{bmatrix}^T \in \mathbb{R}^3$ is the total thrusts which are defined by the control inputs $\mathbf{u}_i(t-\tau)=\mathbf{R_i}(\mathbf{q_i}(t)) \mathbf{F_i}(t-\tau)\in \mathbb{R}^3$ (roll, pitch, and the total thrust) for each UAV. $\mathbf{G} \triangleq \begin{bmatrix} 0 & 0 & -9.81 \end{bmatrix}^T$ and $\mathbf{A}\in \mathbb{R}^{3 \times 3}$ represent the gravity, and the drag-coefficients. $q_{\phi,i}^d, q_{\theta,i}^d$ are the desired input values with time constants $\alpha_\phi, \alpha_\theta$, and gains $K_{\phi}, K_{\theta}$, respectively. Finally, the time delays consisting of $d_1$ and $d_2$ are denoted with $\tau$.

\subsubsection{State Estimator}
As in~\cite{sankaranarayanan2023paced}, in order to compensate for the system's time delays, we leverage the estimated position and velocity of each UAV as presented in Eq.~\eqref{eq:est}:
\vspace{-0.1cm}
\begin{subequations}
\label{eq:est}
    \begin{align}
        \mathbf{\widehat{p}_i}(t) &= \mathbf{p}_i(t-\tau) \label{eq:pos_est} \\
        \implies \mathbf{\dot{\widehat{p}}}_i(t) &= \mathbf{\dot{p}}_i(t-\tau) \label{eq:vel_est} \\
        \mathbf{{\widehat{p}}}_i(t + \tau) &= \mathbf{\widehat{p}}_i(t) + \mathbf{\dot{p}}_i(t) \tau \label{eq:pos_est_fut} \\
        \implies \mathbf{\dot{\widehat{p}}}_i(t + \tau) &= \mathbf{\dot{\widehat{p}}}_i(t) + \mathbf{\ddot{p}}_i(t) \tau \label{eq:vel_est_fut_2}
    \end{align}
\end{subequations}
The above information is used to generate the future states of the UAVs for the control inputs, as described in Eq.~\eqref{eq:vel_est_fut_3}:
\vspace{-0.1cm}
\small
\begin{align}
    \mathbf{\dot{\widehat{p}}}_i(t + \tau) &= \mathbf{\widehat{p}}_i(t) + \left ( \frac{1}{m} \mathbf{u}_i(t-\tau) + \mathbf{G} - \mathbf{A}\mathbf{\dot{\widehat{p}}}(t+\tau) \right ) \tau \label{eq:vel_est_fut_3}
\end{align}
\normalsize
\subsubsection{Centralized Controller}
The objective of the controllers is to design a function so that the control inputs $\mathbf{u}_i$ will generate collision-free trajectories to track the desired reference positions. The function considers both the time delays and the state of every agent of its centralized scheme, in order to generate the paths and to penalize deviation from the desired position. The controllers' complexity lies in the specific number of agents that they have to control, the prediction horizon $N$, and the sampling time, $T$, to optimize the control inputs. The reference states, $\mathbf{x}_i^{ref}$ are sampled through the prediction horizon for formulating the cost function, $J_n$, where $n=1,\dots,CNMPC_{num}$ as described in Eq.~\eqref{eq:mpc_cost}:
\vspace{-0.1cm}
\begin{align}
    J_n &= \sum_{j=0}^{N} \sum_{i=1}^{N_a} \{\left (\mathbf{x}^{ref}_{k+j,i} - \mathbf{x}_{k+j|k,i} \right)^T \mathbf{Q}_{x} \left (\mathbf{x}^{ref}_{k+j,i} - \mathbf{x}_{k+j|k,i} \right) \nonumber \\
    & + \left (\mathbf{u}_{k+j|k,i} - \mathbf{u}_{k+j-1|k,i} \right)^T \mathbf{Q}_{\delta u} \left (\mathbf{u}_{k+j|k,i} - \mathbf{u}_{k+j-1|k,i} \right) \nonumber\\
    & + \left ( \mathbf{u}_{k+j|k,i} + \mathbf{G} \right)^T \mathbf{Q}_{u} \left ( \mathbf{u}_{k+j|k,i} + \mathbf{G} \right)\} \label{eq:mpc_cost}
\end{align}
The cost matrices serve to minimize not only state errors ($\mathbf{Q}_x$) but also ensure the smoothness of the control signal by reducing differences between consecutive inputs ($\mathbf{Q}_{\delta u}$). Moreover, they help maintain control inputs at a proximity to hovering mode ($\mathbf{Q}_u$). Our method, while powerful, involves greater computational complexity compared to alternative approaches. This complexity arises from the necessity for a centralized scheme where all agents ($N_a$) share their states. However, this increased computational load is effectively managed by the monitored cloud resources, which facilitate the execution of controllers and the attainment of optimal solutions.

To enforce collision avoidance as a constraint over the cost function, we consider the estimated positions of the UAVs over the prediction horizon. Hence, an agent-to-agent ($l,i$) collision avoidance constraint $C^{l,i}$ is presented in~\eqref{eq:distance}. The constraint becomes an equality when satisfied, enforcing a minimum separation of $r_\mathrm{safe}$ between each agent.

\begin{multline}\label{eq:distance} 
    C^{l,i}(\bm{x}_{k}) 
    {}\coloneqq{} 
   [r_{\mathrm{safe}}^2 - (p_{x,k+j|k,i}-p_{x,k+j|k,i})^2 
    \\ - (p_{y,k+j|k,i}-p_{y,k+j|k,i})^2]_+ 
    {}={} 0
\end{multline}
The CNMPC problems' are solved with the Proximal Averaged Newton-type method for Optimal Control (PANOC)~\cite{small2019aerial} using the Eq.~\eqref{eq:opt_obj}, while the stability analysis of \cite{seisa2024edge} was used.
\begin{subequations}\label{eq:opt_obj}
\vspace{-0.5cm}
    \begin{align}
        \operatorname*{Minimize}_{
            \bm{u}_k, \bm{x}_k
        } \,
        & J(\bm{x}_{k}, \bm{u}_{k}; u_{k-1\mid k})
        \\
        \text{s.t.:}\,&
        x_{k+j+1\mid k} = \zeta(x_{k+j\mid k}, u_{k+j\mid k})\notag\\
        &\quad j=0,\ldots, N-1
        \\
        &u_{\min} \leq u_{k+j\mid k}^{(i)} \leq u_{\max}, j=0,\ldots, N
    \label{eq:nmpc:input_constraints}
        \\
         & C^{l,i}(\bm{x}_{k}) = 0,  j=0,\ldots, N\notag
         \\
         &\quad i, l=1,\ldots, N_a
         \\
         &x_{k{}\mid{}k}^{(i)} = x_k^{(i)}, i=1,\ldots, N_a
    \end{align}
\end{subequations}


\section{Experimental Setup with Simulated Agents}
\label{sec:simulation}
\begin{figure}[ht]
    \centering
    \includegraphics[width=0.8\linewidth]{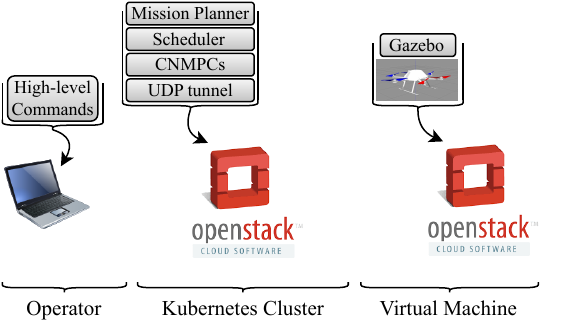}
    \caption{System overview with the real-time cloud test bed operating the scheduling mechanism within a Kubernetes cluster.}
    \label{fig:sim}
\end{figure}

To assess the effectiveness of our proposed approach, we established a test environment within the Ericsson Research real-time cloud infrastructure in Sweden~\cite{ericsson}, which is built upon OpenStack~\cite{rosado2014overview}. This test bed encompasses our Kubernetes cluster and the simulation environment. that is depicted in Fig.~\ref{fig:sim}. For simulating UAVs, we used the simulation ROS package~\cite{furrer2016rotors} for Gazebo, running on a dedicated Virtual Machine within the OpenStack. The primary objective of this research is to evaluate our scheduling mechanism's performance in terms of computational efficiency, communication delays, maintaining precise control with limited tracking error, and its scalability to accommodate a substantial number of agents. \cmmnt{In addition, assessments regarding the uplink, downlink, and CNMPC response time are conducted. These are evaluated with the following experiments.}

\begin{figure}[ht]
    \centering
    \includegraphics[width=0.82\linewidth]{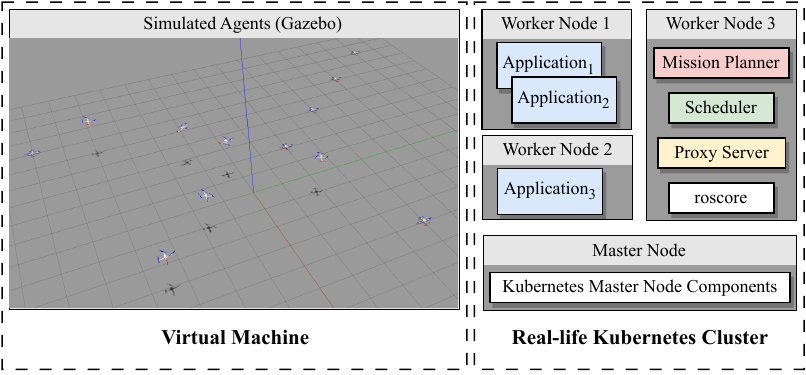}
    \caption{The Kubernetes cluster experimental setup with a snapshot of the external VM that hosts the simulator. Three CNMPCs are contributed to two worker nodes, and control the trajectory of 21 UAVs.}
    \label{fig:setup}
\end{figure}

The proposed Kubernetes cluster comprises one master node and three worker nodes. The mission planner, scheduler, and the UDP tunnel operate continuously on the worker nodes. In contrast, CNMPC pods are dynamically deployed as needed, with their deployment location determined based on resource requirements. The specifications of the Kubernetes cluster and the external machine hosting the simulated agents are presented in Table~\ref{tab:cluster}, while a visual representation of the cluster is depicted in Fig.~\ref{fig:setup}. We utilized Kubernetes version \text{v1.26.1} and Docker version \text{v24.0.5} as the container runtime for this work, with all scheduling mechanism modules hosted within the cluster as Kubernetes pods. Worker node 3, having weaker specifications compared to nodes 1 and 2, hosts the mission planner, scheduler, UDP tunnel, and roscore. In contrast, worker nodes 1 and 2 are dedicated to the controllers. The Kubernetes orchestration seamlessly assigns CNMPCs to worker nodes, taking into account the cluster's availability and the specific resource requirements of each CNMPC, while extra worker nodes can be integrated into the cluster to help manage any overload.

\begin{table}[ht!]
    \centering
    \caption{Kubernetes Cluster and Simulator VM Specifications}
    \begin{tabular}{| c || c | c | c |}
        \hline
        \centering
        & CPU & Memory & Environment \\
        \hline
        \hline
        \centering
        Master Node & 3-core & 2GB & Ubuntu 20.04.6 LTS \\
        \hline
        \centering
        Worker Node 1& 32-core & 460GB & Ubuntu 20.04.6 LTS \\
        \hline
        \centering
        Worker Node 2& 16-core & 32GB & Ubuntu 20.04.6 LTS \\
        \hline
        \centering
        Worker Node 3& 4-core & 8GB & Ubuntu 20.04.6 LTS \\
        \hline
        \centering
        Simulator & 32-core & 480GB & Ubuntu 20.04.6 LTS \\
        \hline
    \end{tabular}
    \label{tab:cluster}
\end{table}

\begin{figure*}
    \centering
    \includegraphics[width=0.78\textwidth]{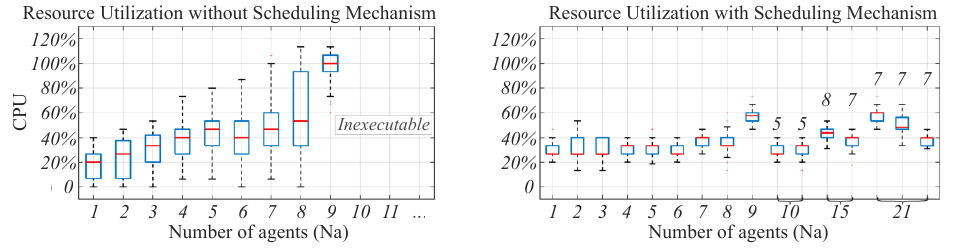}
    \caption{Comparison of resources utilization with (right figure) and without (left figure) scheduling mechanism. The number of agents exceeds 10 (5 on CNMPC$_1$ - Worker node 1, and 5 on CNMPC$_2$ - Worker Node 2), and 15 (8 on CNMPC$_1$ - Worker node 1, and 7 on CNMPC$_2$ - Worker node 2), and scale up to 21 (7 on CNMPC$_1$ - Worker node 1, 7 on CNMPC$_2$ - Worker node 1, and 7 on CNMPC$_3$ - Worker node 2).}
    \label{fig:resource_utilization_scheduling}
    \vspace{-0.5cm}
\end{figure*}

In Fig.~\ref{fig:resource_utilization_scheduling}, we assess the scheduling mechanism and demonstrate resource utilization for a sample of 50 trials. The left figure illustrates resource utilization on a 16-core machine without scheduling, while the right figure shows resource utilization with our proposed mechanism. Without scheduling, resource utilization exponentially rises with an increasing number of agents. In contrast, our approach effectively monitors and maintains resource utilization within desired limits. This figure highlights our approach's capability to scale up the number of agents, overcoming limitations seen in traditional centralized approaches.

\begin{figure}[ht]
    \centering
    \includegraphics[width=0.9\linewidth]{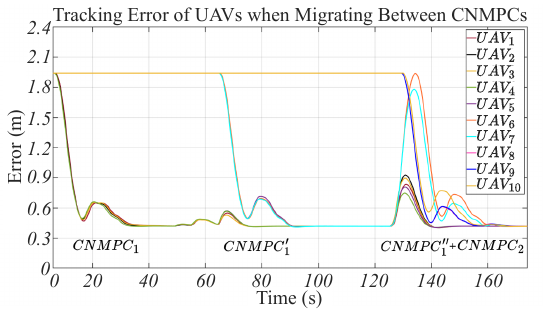}
    \caption{Tracking error of ten UAVs when transitioning between CNMPCs.}
    \label{fig:tracking_error_scaling}
\end{figure}

In Fig.~\ref{fig:tracking_error_scaling}, we illustrate the tracking error of UAVs as they transition between CNMPCs, and we highlight the scaling and migrating capability of our approach. Initially, the mission planner provides control for four UAVs through (CNMPC$_1$). At $t=65$ seconds, three additional UAVs join the system, prompting the scheduler to create a new controller (CNMPC$_1'$) to manage all the robots. When the new CNMPC is deployed, all agents migrate to it. By $t=125$ seconds, three more UAVs join. In response, the scheduler initiates the deployment of two controllers (CNMPC$_1''$ and CNMPC$_2$) to facilitate trajectory control for all robots. UAV$_{6}$ and UAV$_{7}$ must adjust their positions, incurring a tracking error increase (their new set point is far from the current one), to join CNMPC$_{2}$ along with the new additions, while the remaining five UAVs are controlled by the newly generated CNMPC$_{1}''$.

To enable remote cloud-based control for all the agents, we need to ensure bounded time delays in the communication link, regardless of the number of agents.\cmmnt{The succeeding results elaborate on the uplink and downlink delays and examine the relationship between CPU load and CNMPC processing time.} Given that the only messages transmitted between the agents and the cloud are odometry states and control commands\cmmnt{(with \texttt{sizeof}($x_{i}(k)$), \texttt{sizeof}($u_{i}(k)$) $\in (0, 1]$ (Kilobyte))}, the channel's bandwidth can effectively manage the communication load.

\cmmnt{
\begin{figure*}[t]
    \centering
    \includegraphics[width=0.92\textwidth]{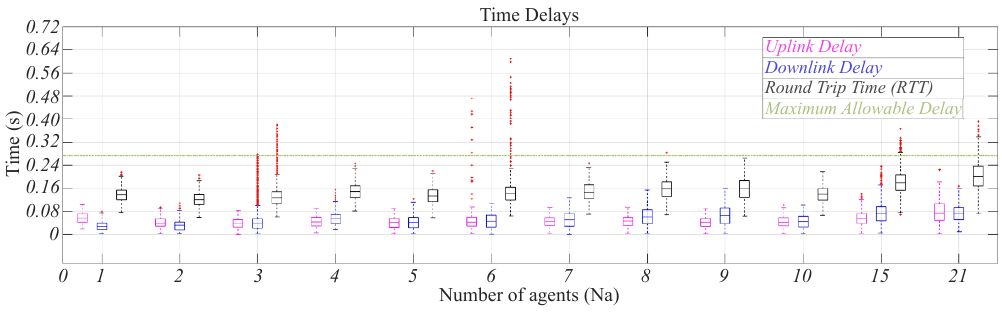}
    \caption{Illustration of uplink delay, downlink delay, and round trip time, along with the maximum allowable delay.}
    \label{fig:delays}
\end{figure*}
}
\cmmnt{
In Fig.~\ref{fig:delays}, the uplink delay, $\tau_u \in \mathbb{R}^+$ ($\tau_u = d_1$) and downlink delay, $\tau_d \in \mathbb{R}^+$ ($\tau_d = d_2$) delays, are depicted, along with the Round Trip Time (RRT), $\tau_{rrt} \in \mathbb{R}^+$, for the closed-loop control system. The RRT is expressed as $\tau_{rrt} = \tau_u + \tau_d + \tau_p$, where, $\tau_p$ represents the processing time required for the CNMPCs to generate feasible control solutions. Consistent with our findings in previous work\cite{seisa2024edge}, system stability requires that $\tau_{rrt}$ not exceed the maximum allowable value, $\tau_{max}$, enforced by the condition in Eq.~\ref{eq:delay}.}

\begin{figure}[ht]
    \centering
    \includegraphics[width=.90\linewidth]{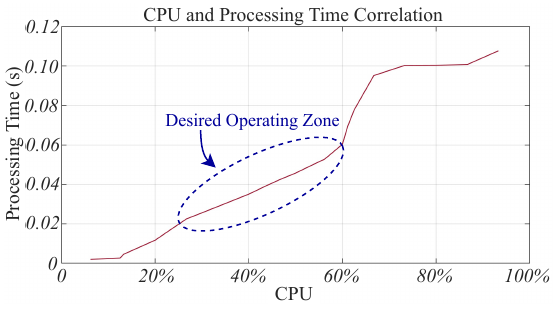}
    \caption{Correlation between the CPU and the processing time along with the target CPU utilization range represented with the blue zone.}
    \label{fig:cpu}
\end{figure}

As we monitor the resources on the real-time cloud, we can minimize $\tau_p$, which is the processing time required for the CNMPCs to generate feasible control solutions, whenever it is needed, as shown in Fig.~\ref{fig:cpu}. Therefore,  by applying a sliding window average technique detailed in~\cite{10228114}, we maintain $\tau_{rrt} = \tau_u + \tau_d + \tau_p$ ($\tau_{rrt}, \tau_u, \tau_d \in \mathbb{R}^+$ are the round trip time, the uplink and downlink delays, respectively) within acceptable limits, independent of the number of agents involved.

\vspace{-0.5cm}
\begin{align}
    \tau_{rrt} &\leq \tau_{max} \label{eq:delay}
\end{align}

Fig.~\ref{fig:cpu} demonstrates the relationship between CPU utilization and the average processing time of the CNMPCs, as given in Eq.~\ref{eq:relation}. The data indicate an increase in processing time associated with higher CPU usage. To maintain processing times within desired limits, CPU usage should be kept within the designated blue zone. This is achieved by monitoring of cloud resources, as defined by Eq.~\eqref{eq:resources_cpu}~\eqref{eq:resources_mem}, which inform the resource scheduling outcomes illustrated in Fig.~\ref{fig:resource_utilization_scheduling}.
\vspace{-0.01cm}
\begin{align}
    \tau_{p}^{avg} &= f(CPU) \label{eq:relation}
\end{align}

Finally, although rule-based scheduling mechanisms are common in cloud computing environments, and allow for the predetermination of rules or policies as discussed in~\cite{murad2022review}, they fall short in ensuring application-specific requirements. In contrast to~\cite{murad2022review}, our work experimentally demonstrates that our scheduling approach not only meets the desired CNMPCs performance while respecting predefined requirements, but also dynamically interacts with the applications. This enables the modification of their operating points and requirements to ensure optimal performance.


\section{Conclusions and Future Developments}
\label{sec:conclusions}
In this work, we have presented a novel approach to address the challenges of deploying complex robotic software in large-scale systems, i.e., CNMPCs for multi-agent systems. Our system leverages cloud computing and intelligent scheduling to provide a dynamic and scalable solution for the robotic application. We highlighted the advantages of our system, particularly its ability to handle a variable number of robots. Through experimental tests, we have demonstrated the effectiveness and performance of our system, especially in scenarios where the number of robots is subject to change. Our approach not only optimizes resource utilization but also offers flexibility and adaptability.

Extending our system to real-world robotic deployments is a crucial next step. Testing our approach in diverse environments and with various robot types will provide valuable insights and validation. While the proposed mechanism is evaluated in centralized control schemes, it can be applied to distributed systems, and application agnostic scenarios. Finally, more advanced optimization algorithms for intelligent scheduling can further improve resource allocation and task execution efficiency.

%
\section{Acknowledgement}
\label{sec:acknowledgement}
The authors would like to express their gratitude to the Ericsson Cloud Research team, in Lund, Sweden.
\bibliographystyle{./IEEEtranBST/IEEEtran}
\bibliography{./IEEEtranBST/IEEEabrv, references}

\end{document}